\documentstyle[12pt]{article}
\newcommand{\di}{\displaystyle}
\newcommand{\ga}{\gamma}
\newcommand{\si}{\sigma}
\newcommand{\eV}{\;\mbox{\rm eV}}
\newcommand{\GeV}{\;\mbox{\rm GeV}}

\newcommand{\cm}{\; \mbox{\rm cm}}
\newcommand{\met}{\; \mbox{\rm m}}

\newcommand{\pc}{\; \mbox{\rm pc}}

\begin{document}

\setlength{\parskip}{0.45cm} \setlength{\baselineskip}{0.75cm}
%XXXXXXXXXXXXXXXXXXXXXXXXXXXXXXXXXXXXXX
%
%SETTINGS FOR PREPRINT-SPACED VERSION
%setlength{\parskip}{0.45cm}
%setlength{\baselineskip}{0.75cm}
%
% SETTINGS FOR DOUBLE - SPACED VERSION
%\setlength{\parskip}{0.65cm}
%\setlength{\baselineskip}{0.95cm}%
%XXXXXXXXXXXXXXXXXXXXXXXXXXXXXXXXXXXXXX
\begin{titlepage}
\setlength{\parskip}{0.25cm} \setlength{\baselineskip}{0.25cm}
\begin{flushright}
DO-TH 02/11\\
\vspace{0.2cm}
%hep--ph/0103137\\
%\vspace{0.2cm}
%November 2001
\end{flushright}
\vspace{1.0cm}
\begin{center}
\Large
{\bf Antineutrino-neutrino and antineutrino-electron}\\
\Large
{\bf resonant annihilation through rho}\\
\Large {\bf and other vector mesons} \vspace{1.5cm}

\large
Emmanuel A.\ Paschos and Olga Lalakulich\\
\vspace{1.0cm}

\normalsize
{\it Universit\"{a}t Dortmund, Institut f\"{u}r Physik,}\\
{\it D-44221 Dortmund, Germany} \\
\vspace{0.5cm}

\vspace{1.5cm}
\end{center}

\begin{abstract}
The $\bar{\nu}e^-\to\pi^-\pi^0$ and $\nu e^+\to \pi^+\pi^0$ reactions
have a resonant structure whenever the energy of the $s$--channel
equals the mass of a $J^P=1^-$ vector meson. The resonant cross
sections are of the order $10^{-38}$ cm$^2$ and correspond to cosmic
ray antineutrino energies in the range 0.20 to 2.00 TeV. A similar
structure occurs in the annihilation of cosmic ray antineutrinos with a
relic (background) neutrino.  For $m_{\nu}\leq 10^{-3} \eV$ the
resonant energy is above GZK limit and their decay products include
multiple $\gamma$--rays.  Possible detection schemes are discussed,
especially those which rely on Cherenkov radiation.
\end{abstract}
\end{titlepage}

\input FEYNMAN

%Sect.1
\section{Neutrino--initiated reactions in water.}
The first-generation neutrino telescopes (Lake Baikal, AMANDA) are in
operation and the next-generation telescopes are under construction.
The main purpose of these experiments is to detect cosmic neutrinos
(and/or antineutrinos) with energy in TeV-EeV region, which are believed
to be coming to the earth from gamma-ray bursts and active galactic
nuclei. For their detection the following neutrino interactions in
water are usually considered.

1. Charged--current induced neutrino-nucleon reactions
\cite{mohapbook1991}
%Eq.(1)
\begin{equation}
\nu_l \, N \to l^- \, X   \qquad
\si (\nu N) \approx 0.67 \cdot 10^{-38}
  \left(\frac{E_\nu}{1\GeV}\right) \cm^2\, ,
\end{equation}
and antineutrino interactions
%Eq.(2)
\begin{equation}
\bar\nu_l \, N \to l^+ \, X  \qquad
\si (\bar\nu N) \approx 0.34 \cdot 10^{-38}
 \left(\frac{E_\nu}{1\GeV}\right) \cm^2\, .
\end{equation}

Here the cross sections are averaged over proton and neutron; the
reactions are sensitive to all neutrino flavours. The above
cross-sections are valid at energies in the range $m_l^2/2m_N \ll E_\nu
\ll m_W^2/2m_N$. It is these reactions that the experiments rely upon
for detection of atmospheric neutrinos. The high-energy recoil lepton
produces Cherenkov light and is detected as the so-called one-ring
event. These cross sections are the largest among the reactions
available in the experiments. The same reactions on nuclei are
calculated as the incoherent sum of protons and neutrons with the
nuclear corrections being small \cite{gandhi96}.

2. Neutral--current induced neutrino-nucleus interactions have cross
sections three times smaller and are not considered, because it is hard
to identify the final products.

3. Elastic neutrino-electron and antineutrino-electron
scattering
\[
\nu_l e^- \to \nu_l e^-  \qquad
 \bar\nu_l e^- \to \bar\nu_l e^-
\]
reactions are also sensitive to all three flavours of neutrinos. The
high-energy recoil electron is observed through the produced Cherenkov
light. The corresponding cross-sections (for $m_l^2/2m_e \ll E_\nu \ll
m_W^2/2m_e= 6.4\cdot 10^9 \GeV)$ are as follows \cite{mohapbook1991}:

\begin{eqnarray} \di
\sigma(\nu_e e^- \to \nu_e e^-) & = &
  \frac{G_{\mathrm{F}}^2m_eE_\nu}{2\pi}\left[
    (2\sin^2\theta_{\mathrm{W}}+1)^2+\frac43\sin^4\theta_{\mathrm{W}} \right]
\nonumber \\
   & \approx & 0.9\cdot 10^{-41}\cdot\left(\frac{E_\nu}{1\GeV}\right)\cm^2
\nonumber \\[2mm]    \di
\sigma(\bar\nu_e e^- \to \bar\nu_e e^-) & = &
  \frac{G_{\mathrm{F}}^2m_eE_\nu}{2\pi}\left[
   \frac13(2\sin^2\theta_{\mathrm{W}}+1)^2+\frac43\sin^4\theta_{\mathrm{W}} \right]
\nonumber \\  &  \approx & 0.378\cdot
10^{-41}\cdot\left(\frac{E_\nu}{1\GeV}\right)\cm^2
\nonumber \\[2mm]   \di
\sigma(\nu_{\mu(\tau)} e^- \to \nu_{\mu(\tau)} e^-)
 & = & \frac{G_{\mathrm{F}}^2m_eE_\nu}{2\pi}\left[
   (2\sin^2\theta_{\mathrm{W}}-1)^2+\frac43\sin^4\theta_{\mathrm{W}} \right]
\nonumber \\    & \approx & 0.15\cdot
10^{-41}\cdot\left(\frac{E_\nu}{1\GeV}\right)\cm^2
\nonumber \\[2mm]   \di
\sigma(\bar\nu_{\mu(\tau)} e^- \to \bar\nu_{\mu(\tau)}
  e^-) & = & \frac{G_{\mathrm{F}}^2m_eE_\nu}{2\pi}\left[
   \frac13(2\sin^2\theta_{\mathrm{W}}-1)^2+\frac43\sin^4\theta_{\mathrm{W}} \right]
\nonumber \\  &  \approx & 0.14\cdot
10^{-41}\cdot\left(\frac{E_\nu}{1\GeV}\right)\cm^2 \nonumber
\end{eqnarray}

The purpose of this paper is to call attention to $\bar\nu_l l^-$ (and
$\nu_l l^+$) resonant annihilation mediated by a negatively-- (and
positively) charged vector meson $V$ with $J^P=1^-$. Among the charged
leptons only electrons are contained in targets on the earth. Thus only
the reaction $\bar\nu_e e^-$, which is sensitive to electron
antineutrinos in energy range from about $\sim 200\GeV$ to $\sim 2000
\GeV$, is of particular interest for experimental physics.

\begin{figure}[htb]
\begin{center}
\begin{picture}(30000,6000)
\drawline\fermion[\NW\REG](3000,3000)[3700]  \put(0,4400){$\bar\nu_e$}
\drawline\fermion[\SW\REG](3000,3000)[3700]  \put(0,700) {$e^-$}
\drawline\photon[\E\REG](3000,3000)[4]    \put(4000,3500) {$W^-$}
\put(\photonfrontx,\photonfronty){\circle*{200}}
\put(\photonbackx,\photonbacky){\circle*{500}}
\drawline\photon[\E\REG](\photonbackx,\photonbacky)[4]
\put(9000,3500){$V^-$} \put(\photonbackx,\photonbacky){\circle*{200}}
\drawline\scalar[\NE\REG](\photonbackx,\photonbacky)[2]
\drawline\scalar[\SE\REG](\scalarfrontx,\scalarfronty)[2]
\put(17000,3000){$\equiv$}
\drawline\fermion[\NW\REG](23000,3000)[3700]
          \put(\fermionbackx,4000){$\bar\nu_e$}
                      \put(21400,5000){$p_{\bar\nu}^\mu$}
\drawline\fermion[\SW\REG](23000,3000)[3700]
           \put(\fermionbackx,1100) {$e^-$}  \put(21000,0){$p_e^\mu$}
\drawline\photon[\E\REG](\fermionfrontx,\fermionfronty)[5]
                \put(25000,3500){$V^-$}
\put(25000,2000){$q^\mu$}
\put(\photonfrontx,\photonfronty){\circle{200}}
\put(\photonfrontx,\photonfronty){\circle{400}}
\put(\photonfrontx,\photonfronty){\circle{600}}
%\put(\photonbackx,\photonbacky){\circle*{200}}
%                        \put(24600,9000){$\nu_\tau$}
\put(\photonbackx,\photonbacky){\circle*{200}}
\drawline\scalar[\NE\REG](\photonbackx,\photonbacky)[2]
\put(\scalarbackx,4500){$p_1^\mu$}
\drawline\scalar[\SE\REG](\scalarfrontx,\scalarfronty)[2]
\put(\scalarbackx,1500){$p_2^\mu$}
\end{picture}
\end{center}
\caption{Antineutrino-electron resonant annihilation
through charged $V$-meson with $J^P=1^-$.}
\label{barnue}
\end{figure}
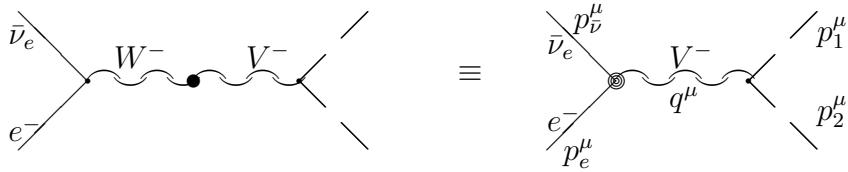
\noindent The main contribution to the $\bar\nu_e e^-$ annihilation
comes from the resonance in the intermediate state, which corresponds
to neutrino energies $E_{\nu(res)}=m_V^2/2m_e$. The calculation of this
process becomes obvious when we keep in mind calculations with the
vector meson dominance.

A similar process of high energy neutrino (antineutrino) annihilation
takes place with the relic background antineutrino (neutrino) mediated
by a neutral vector meson $V^0$. It would be remarkable if this process
could serve as a means for the detection of the relic background
neutrinos and we shall discuss it in section 3.

%Sect.2
\section{Antineutrino-electron resonant annihilation}
The lightest $J^P=1^-$-meson is the $\rho^-(770)$-meson and we start
our discussion with this particle. Recent observations of the $W^- \to
\rho^-$ transition, which clearly determine the $W-\rho$ coupling, come
from the $\tau-$decays $\tau^-\to \nu_\tau \pi^- \pi^0$ \cite{pdg}.

\begin{figure}[htb]
\begin{center}
\begin{picture}(30000,10000)
\drawline\fermion[\E\REG](0,7000)[4000] \put(0,7400){$\tau^-$}
\put(\fermionbackx,\fermionbacky){\circle*{200}}
\drawline\photon[\SE\REG](\fermionbackx,\fermionbacky)[3]
           \put(5200,6100){$W^-$}
\drawline\fermion[\NE\REG](\fermionbackx,\fermionbacky)[3000]
           \put(4600,9000){$\nu_\tau$}
\put(\photonbackx,\photonbacky){\circle*{500}}
\drawline\photon[\SE\REG](\photonbackx,\photonbacky)[3]
                    \put(7300,4700){$\rho^-$}
\put(\photonbackx,\photonbacky){\circle*{200}}
\drawline\scalar[\NE\REG](\photonbackx,\photonbacky)[2]
            \put(\scalarbackx,4600){$\pi^-$}
\drawline\scalar[\SE\REG](\scalarfrontx,\scalarfronty)[2]
                  \put(\scalarbackx,800){$\pi^0$}
\put(17000,5000){$\equiv$} \drawline\fermion[\E\REG](20000,7000)[4000]
\put(20000,7400){$\tau^-$}
%\put(\fermionbackx,\fermionbacky){\circle*{200}}
\drawline\fermion[\NE\REG](\fermionbackx,\fermionbacky)[3000]
\put(24600,9000){$\nu_\tau$}
\drawline\photon[\SE\REG](\fermionfrontx,\fermionfronty)[4]
\put(26000,5600){$\rho^-$} \put(\photonfrontx,\photonfronty){\circle{200}}
\put(\photonfrontx,\photonfronty){\circle{400}}
\put(\photonfrontx,\photonfronty){\circle{600}}
\put(\photonbackx,\photonbacky){\circle*{200}}
\drawline\scalar[\NE\REG](\photonbackx,\photonbacky)[2]
            \put(\scalarbackx,5700){$\pi^-$}
\drawline\scalar[\SE\REG](\scalarfrontx,\scalarfronty)[2]
                  \put(\scalarbackx,2100){$\pi^0$}
\end{picture}
\end{center}
\caption{Decay $\tau^-\to \nu_\tau \pi^- \pi^0$ through
the $\rho^-$meson.}
\end{figure}
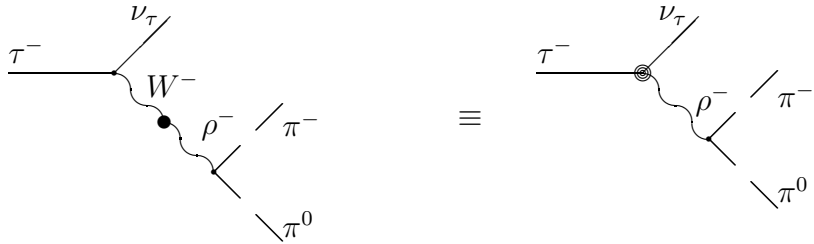

The process $\tau^- \to \nu_\tau \rho^- \to \nu_\tau \pi^- \pi^0$
contributes $(25.4\pm 0.14)\%$ to the total $\tau-$width, while the
process without $\rho^-$ in the intermediate state contributes at most
$(0.3\pm 0.32)\%$. It is clear, that pions in the final state are
coming completely from the $\rho-$meson decay, since the $\pi^-\pi^0$
decay channel contributes $\sim 100\%$ \cite{pdg} to the total width.

We use the central values of the data to extract the phenomenological
$\tau\rho\nu_\tau$ coupling. Indeed, following the discussion by Okun
\cite{okunbook}, we write the matrix element of the
$\tau^-\to\nu_\tau\rho^-$ decay as
%Eq.(3)
\begin{equation}
M=\frac{G_{\mathrm{F}}}{\sqrt{2}} g_\rho \cos\theta_c
  \varphi_\alpha^{(\rho)}
\bar{u_\nu} \gamma^\alpha (1-\gamma_5) u_\tau,
\end{equation}
where $\theta_c$ is the Cabbibbo angle, $\varphi_\alpha^{(\rho)}$ is
$\rho-$meson wave function. The result of the width calculation reads
%Eq.(4)
\begin{equation}
(0.254-0.003)\cdot \Gamma_{\tau(tot)}
=\frac{G^2}{16\pi} g_\rho^2\cos^2\theta_c
\frac{m_\tau^3}{m_\rho^2}\left(1-
 \frac{m_\rho^2}{m_\tau^2}\right)^2
  \left(1+\frac{2m_\rho^2}{m_\tau^2}\right),
\end{equation}
which for $m_\tau=1.777\GeV$,
$\Gamma_{\tau(tot)}=2.26\cdot 10^{-12}\GeV$
and $m_\rho=0.7665\GeV$ gives
%Eq.(5)
\begin{equation}
g_\rho^2\cos^2\theta_c=0.02435\, .
\label{g2rhocos2}
\end{equation}
We subtracted in Eq.\ (4) a $0.3\%$ contribution for the non--resonant
background. Recalling that $\cos\theta_c=0.974$, our result is in
agreement with the values obtained from indirect estimates, mentioned
in \cite{okunbook}. This value for $g_\rho^2\cos^2\theta_c$ is obtained
for $\rho$ on--the--mass--shell. For virtual $\rho-$meson, strictly
speaking, $g_\rho^2\cos^2\theta_c$ can vary with center-of-mass energy
$\sqrt{s}$. However, up to now all calculations within the vector meson
dominance framework (which are very extensive, for example, for
neutrino--nucleus scattering \cite{pumplin90}) never take into account
this dependence. Since in our case the dominant contribution to
annihilation process occurs at $\sqrt{s}=m_\rho\pm \Gamma_\rho/2$ (i.e.
$\rho$ is on-mass-shell or almost on-mass-shell), we neglect the
variation of the coupling constant with $\sqrt{s}$.

Another quantity we need to know for the calculation is the
$\rho\pi\pi$ vertex, which is obtained from the $\rho-$meson width. The
vertex, according to the effective gauge model of hadron interactions,
is written as $if_{\rho\pi\pi}(p_1^\mu-p_2^\mu)$. The width in the
$\rho-$rest frame is
%Eq.(6)
\begin{equation}
\Gamma_{(0)}(\rho\to\pi^-\pi^0)
=\frac23\frac{f_{\rho\pi\pi}^2}{4\pi}
  \frac{|p_1|^3}{m_\rho^2},
   \qquad |p_1|=\frac{\sqrt{m_\rho^2-4\cdot m_\pi^2}}{2}
\end{equation}
For $\Gamma_{\rho(0)}=0.15\GeV$, $m_\pi=0.137\GeV$ one obtains from
the width
%Eq.(7)
\begin{equation}
\frac{f_{\rho\pi\pi}^2}{4\pi}=2.896.
\label{frhopipi}
\end{equation}

Now it is straightforward to write down the amplitude for the  resonant
antineutrino--electron annihilation process $\bar\nu_e e^-\to\rho^-\to
\pi^-\pi^0$
%Eq.(8)
\begin{equation}
M=\frac{G_{\mathrm{F}}}{\sqrt{2}}g_\rho\cos\theta_c \bar u_\nu
    \gamma^\alpha(1-\gamma^5) u_e  \frac{-g_{\alpha\beta}
+\frac{q_\alpha q_\beta}{m_\rho^2}}{q^2-m_\rho^2+{\mathrm{i}}
    \Gamma_\rho m_\rho}
      (if_{\rho\pi\pi}) (p_1^\beta - p_2^\beta),
\end{equation}
where $q=p_e+p_{\bar\nu}$,
and to calculate the cross section
%Eq.(9)
\begin{equation}
\si=\frac{G_{\mathrm{F}}^2}3\frac{g_{\rho\pi\pi}^2}{4\pi}
   f_\rho^2\cos^2\theta_c
\frac{2m_e E_\nu}{(2m_e E_\nu-m_\rho^2)^2+2m_e E_\nu \cdot
    \Gamma_{\rho(0)}^2}
\left( 1-\frac{4m_\pi^2}{2m_e E_\nu}\right)^{3/2}
\end{equation}
At resonance the cross-section is
%Eq.(10)
\begin{equation}
\si_{res}(\bar\nu_e e^- \to
   \rho^- \to\pi^-\pi^0)=4.4\cdot 10^{-38}\cm^2\,.
\end{equation}
With this cross-section one can easily estimate the expected number of
events in a water detector of volume $V$
%Eq.(11)
\begin{equation}
N=\int d\Omega \,  N_A \, \rho_{H_2 O} \,  V \,
  \int\limits_{E_{\nu(min)}}^{E_{\nu(max)}}
   j(E_\nu,\theta) \si(E_\nu) d E_\nu
\end{equation}
The resonant neutrino energy for this process is
$E_{\nu(res)}=m_\rho^2/2m_e=580\GeV$. The main contribution to the
expected number of events comes from the energy interval
$E_{min}=(m_\rho-\Gamma_\rho)^2/2m_e=370\GeV$ to
$E_{max}=(m_\rho+\Gamma_\rho)^2/2m_e=830\GeV$. As the lowest estimate
of the neutrino flux at these energies we can take the atmospheric
neutrino flux ${d\,j(E_\nu)}/{d E_\nu}(\theta,E_\nu)$, calculated by
Volkova \cite{volkova80}. In what follows we neglect the dependence on
the incident angle $\theta$ and make calculations for the vertical
flux.

In the energy region $10^2\GeV-10^6\GeV$, two analytic
formulas are given
in Ref. \cite{volkova80}. The more accurate one
($E_\nu$ must be given in GeV)
%Eq.(12)
\begin{eqnarray}
j_{\bar\nu_e}(E_\nu) =  \frac12\cdot 2.4\cdot 10^{-3}\cdot
  E_\nu^{-2.69}
& \left(
  \frac{0.05}{1+1.5E_\nu/8760}
    +\frac{0.185}{1+1.5E_\nu/1890}  \right. \nonumber
\\[3mm]
    & +  \left. \frac{11.4 E_\nu^{0.083-0.215\log E_\nu}}
      {1+1.21E_\nu/1190}
\right)
\end{eqnarray}
underestimates the flux in comparison to the values presented in the
table of the same reference; the less accurate one
%%Eq.(13)
\begin{equation}
j_{\bar\nu_e}(E_\nu)=\frac12 \cdot 1.26 \cdot
  E_\nu^{-3.69}             %8\cdot 10^{-5}\cdot (8760+3.7\cdot 1890)
\end{equation}
overestimates the flux. Another point of ambiguity is related to the
values of $E_{min}$ and $E_{max}$, a convenient definition for narrow
resonances is to take one width around the resonance. Since the width
of $\rho-$meson is about $20\%$ of its mass and since we know the
behaviour of the cross section and the flux in the whole energy region,
the integration should be done rather over two- or even three-width
intervals, instead of one--width only. The factor $[1\div 3]$ in Eq.\
(\ref{numberofevents}) reflects these ambiguities. Finally, the number
of events obtained is
%Eq.(14)
\begin{equation}
N=20\cdot [1\div 3] \cdot \frac{V}{10^9 \met^3}\cdot
   \frac{\int d\Omega}{2\pi}
    \left( \frac{events}{year}\right)
\label{numberofevents}
\end{equation}

The process under consideration will provide a distinct signature in
the detector. For the resonant netrino energy $E_\nu=580\GeV$ the
characteristic energy of outgoing pions is $E_{\pi^-}\sim E_{\pi^0}\sim
E_\nu/2 \sim 290\GeV$. The characteristic longitudial momentum of each
pion has the same value, but the characteristic transverse momentum is
about $\sqrt{2m_e E_\nu}/2\sim 0.38\GeV$. Thus, the two pions move
nearly collinearly with the angle between them being
$\delta=\sqrt{2m_e/E_\nu}\approx 1.8\cdot 10^{-3}$ rad.

Charge pions with such energy do not decay within the detector (the
mean free path is about 15 kilometers), but can produce a nuclear
shower. If not, the pion is detected by the Cherenkov light produced in
a Cherenkov ring of almost maximal radius (the pion velocity is
$\beta=(1-10^{-7})$ to be compared to the accuracy $\delta \beta/\beta
\sim 10^{-4}-10^{-5}$ \cite{pdg} that can be resolved nowadays at
best). The half-angle $\theta_{cher}$ of the Cherenkov cone is
%Eq.(15)
\begin{equation}
\theta_{\mathrm{Cher}}(\pi^-)=\arccos\frac{1}{n\beta},
\end{equation}
where $n=1.35$ is index of refraction for water.

The neutral pion decays on two photons, each photon having approximate
energy of $145\GeV$ and producing an electromagnetic shower in the
water. Since the energies of photons are high $(>0.4\GeV)$, they are
not distinguishable by the detector. Since the initial photon energy is
shared between many produced electrons, the energy of every electron is
of a few MeV, the velocity is less than $\approx 0.95$, so the
half-angle $\theta_{\mathrm{Cher}}$ of Cherenkov cone is smaller than
that for the $\pi^-$. Thus, the final event from the $\rho^-$ decay is
seen as two nearly collinear Cherenkov rings and can be easily
separated from the elastic scattering initiated one--ring events.

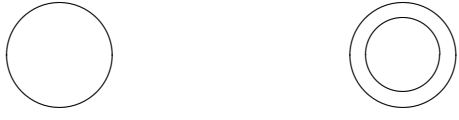
\begin{figure}[htb]
\begin{center}
\begin{picture}(25000,4000)
\put(2000,2000){\circle{4000}}
\put(15000,2000){\circle{3000}}
\put(15000,2000){\circle{4000}}
\end{picture}
\caption{One--ring and two--collinear-rings Cherenkov
detector events.}
\end{center}
\end{figure}

The next vector meson to serve as intermediate state for the process in
Fig.\ \ref{barnue}, is $K^{*-}(872)$ with $m_{K^*}=0.892\GeV$. Its
total width $\Gamma_{K^*}=0.0508\GeV$ originates completely from the
channel $K^*\to K \pi$, so the resonant annihilation produces a pion
and a kaon in the final state: $\bar\nu_e e^- \to K^{*-}\to (K\pi)^-$.

The coupling for the $W-K^*$ transition also appears in $\tau-$decays.
Indeed, the partial width of $\tau^- \to \nu_\tau K^*{}^- \to \nu_\tau
\pi K$ equals $(0.9\pm 0.4)\%$ of the total $\tau-$width, while without
$K^*{}^-$ in the intermediate state the partial width is less than
$<0.17\%$. This gives
%Eq.(16)
\begin{equation}
g_{K^*}^2\sin^2\theta_c=1.124\cdot 10^{-3},
\label{g2kstarsin2}
\end{equation}
which is one order smaller than for the $\rho-$meson.

To derive the $K^*K\pi$ vertex, we introduce for the
$K^*$ decay the coupling
$if_{K^* K \pi}(p_1^\mu-p_2^\mu)$
and  obtain
%Eq.(17)
\begin{equation}
\frac{f_{K^* K \pi}^2}{4\pi}=2.554\, .
\end{equation}
The neutrino-antineutrino annihilation
$\bar\nu_e e^- \to K^{*-}\to (K\pi)^-$
cross section in this case is
%Eq.(18)
\begin{equation}
\begin{array}{l} \di
\si=\frac{G_{\mathrm{F}}^2}3
 \frac{g_{K^* K \pi}^2}{4\pi}f_{K^*}^2\sin^2\theta_c
  \frac{2m_e E_\nu}
   {(2m_e E_\nu-m_{K^*}^2)^2+2m_e E_\nu \cdot
    \Gamma_{K^*(0)}^2} \times
\\[6mm] \di \hspace*{60mm} \times
\left( 1-\frac{m_\pi^2+m_K^2}{m_e E_\nu}+
  \frac{(m_K^2-m_\pi^2)^2}{m_e^2 E_\nu^2} \right)^{3/2}
\end{array}
\end{equation}

The corresponding resonant neutrino energy is $E_{\nu(res)}=780\GeV$.
At resonance the cross section equals $1.0\cdot 10^{-38}\cm^2$ (which
is 4 times smaller than for the $\rho^-$meson). The final states $\pi^0
K^-$ and $\pi^- K^0$ occur with equal probabilities. In the former case
the charged kaon and neutral pion can be observed as two nearly
collinear rings (the same signature as if the $\rho^-$ was in the
intermediate state). This increases the number of events in the
detector by  about $5\%$. In the latter case the neutral kaon can be
either $K_L$ or $K_S$. The long--lived kaon $K_L$ does not decay within
the detector (the mean free path is about 12 kilometers), so only the
charged pion can be observed as one--ring event, which is impossible to
distinguish from the muon resulting from the deep inelastic scattering.
The short--lived kaon $K_S$ with probability $31.4\%$ decays (the mean
free path is 20 meters) on 2 neutral pions and the event as a whole is
again seen as two nearly collinear rings. This again increases the
number of events by about $1.5\%$. With the probability $68.6\%$ $K_S$
decays on two charged pions; in this case the final state is 3 charged
pions which are seen as one-ring event.

Besides $\rho(770)$ and $K^*(892)$, several other charged $J^P=1^-$
mesons are known: $\rho(1700)$, $K^*(1680)$, $\rho(1450)$, $K^*(1410)$.
They can also be intermediate states of antineutrino-electron
annihilation, but up to now there are no experimental data, which we
can use in order to calculate their transition probability to
$W^-$meson. However, it is not unnatural to suppose that their
contribution to the antineutrino--electron annihilation cross section
is of order $10^{-38}\cm^2$.  Thus cross sections of this magnitude
also appear in the interactions of cosmic neutrinos with energies up to
$3-4$ TeV.

Thus, the resonant neutrino annihilation events, when detected with
neutrino telescopes, provide an estimate for the number of cosmic
electron--type antineutrinos. Combined together with the total cross
section events, it provides an estimate of the neutrino fluxes for
electron--type and muon--type neutrinos separately. The importance of
estimating the cosmic neutrino fluxes of various species and of
antineutrinos with high enough accuracy is obvious because they are
relevant for understanding their origin, i.e.\ how and where they are
produced and the ratios of the fluxes when they arrive on the earth.

Finally we mention the Glashow resonance \cite{glashow}, which is
mediated by the $W$--boson $\bar{\nu}_e e^-\to W\to \bar{\nu}_e e^--,\,
\bar{\nu}_{\mu}\mu^-,\, \ldots\,$. The resonant reaction occurs for
$E_{\nu}\approx 6\cdot 4\times 10^{15}\eV$ and has a large cross
section $\approx 10^{-32}$ cm$^2$.

%Sect.3
\section{Antineutrino-neutrino annihilation.}
As mentioned earlier, for experiments on the earth only the electrons
are possible leptonic targets. On a larger scale, like regions of the
universe, there are also relic neutrinos and antineutrinos which are
part of the cosmic background radiation and have not been detected yet.
Their density is estimated to be 54/cm$^3$ for each species having a
temperature of $1.9^o$ K.

A high energy neutrino of each species can annihilate with the
corresponding relic antineutrino and vice versa with neutral vector
mesons forming the intermediate state.  At present we have good
evidence from oscillation experiments that
%Eq.(19)+(20)
\begin{eqnarray}
\Delta m_{\mathrm{atm}}^2 & \sim & 10^{-3} \eV^2
\quad {\rm and}\\
\Delta m_{\mathrm{solar}}^2 & \sim & 10^{-5} \eV^2
\quad {\rm or}\,\,\,{\rm less}\,.
\end{eqnarray}
They allow neutrino masses of the order of $0.01-1\eV$.  These values
are not unique because there is still an arbitrary overall scale. Such
small masses were justifiably neglected when we calculated the neutrino
energy in the previous sections.  The masses must be kept, however,
when the relic neutrinos are considered as targets.

The annihilation of cosmic neutrinos with relic neutrinos with the
Z--boson being an intermediate state was considered twenty years ago
\cite{weiler82,weiler84} and its importance for cosmology and cosmic
ray physics was described recently in \cite{weiler99} and
\cite{ringwald}, where possible implications for neutrino masses are
discussed. More precisely, the resonant cosmic neutrino energy, $E_c$,
for annihilation through the Z--boson is
%Eq.(21)
\begin{equation}
E_c = \frac{M_Z^2}{2m_{\nu}} = 4 \times 10^{21} \eV
\left(\frac{1\eV}{m_{\nu}}\right)
\end{equation}
with $m_{\nu}$ the mass of the lightes neutrino and the energy--averaged
cross section is
%Eq.(22)
\begin{equation}
\sigma(Z)=4 \times 10^{-32}{\rm cm}^2\, .
\end{equation}
The Z--boson created in the reaction decays into several dozens of
secondary particles (baryons, mesons, leptons and neutrinos) known as
``Z--bursts''. A typical ``Z--burst'' contains 2.7 hadrons, 30 photons
and 28 neutrinos. These secondaries form a highly collimated flux, the
half-angle of the the cone being about $2\cdot 10^{-11}$ rad. If the
neutrino--antineutrino annihilation takes place and the following
$Z-$decay occurs within a $20\pc$ zone around the earth and the flux is
directed towards the earth, then the secondaries arrive on the earth
simultaneously. Some of them are able to initiate the multiple air
showers which could be searched for on the earth with air shower
arrays. Therefore, showers of particles which arrive on earth in
coincidence are unique signatures for the neutrino--antineutrino
annihilation, with either the neutrino or antineutrino belonging to the
cosmic background.

For the annihilation process through vector mesons, the resonant neutrino
energy is smaller $E_\nu={M_V^2}/{2m_{\nu}}$. For the $\rho^0$--meson the
relevant reaction
%Eq.(23)
\begin{equation}
\bar{\nu}_{\ell}+\nu_{\ell}\to \rho^0\to \pi^+\pi^-
\end{equation}
occurs at $E_{\rm res}= 3 \times 10^{17}{\eV}
\left({1 \eV}/{m_{\nu}}\right)$.
Comparing this reaction with the annihilation in figure 1, we note that
the kinematics of the two reactions and $f_{\rho\pi\pi}$ are the same.
The couplings $g_{\rho^-}$ (above it was denoted simply as $g_\rho$)
and $g_{\rho^0}$ to the $W^-$ and the $Z^0$--bosons, respectively, are
different. Both vector bosons couple to the isovector part of the weak
vector current and are related by CVC.

The effective Lagrangian for strangeness conserving
processes, relevant to our problem, is given by
%Eq.(24)
\begin{equation}
{\cal L} = \frac{G}{\sqrt{2}}
 \left[ \bar{e}\gamma_{\alpha}(1-\gamma_5)\nu
  \left(J_1^{\alpha} +{\mathrm{i}} J_2^{\alpha}\right)+{\rm h.c.}
   +\bar{\nu}\gamma_\alpha (1-\gamma_5)\nu
    \left(xV_3^{\alpha}- A_3^\alpha + yJ_s^{\alpha}\right)
     \right]
\end{equation}
where $J_i=V_i-A_i$ is one of the isospin components of the usual $V-A$
currents and $J_s$ the isoscalar current.  In the electroweak theory
%Eq.(25)
\begin{equation}
x = 1-2\sin^2\theta_{\mathrm{W}}\quad{\rm and}\quad
 y = -2\sin^2\theta_{\mathrm{W}}
\end{equation}
with $J_s=\frac{1}{\sqrt{3}}V_8$, $\theta_{\mathrm{W}}$ the weak mixing
angle and we neglect strange quark contributions.
%\vspace{5.0cm}
%
%\begin{center}
%Fig.\ 4
%\end{center}

The $\rho$--mesons couple to the $V_1^{\mu}\pm {\mathrm{i}} V_2^{\mu}$
and $V_3^{\mu}$ parts of vector current and their couplings are defined
as
%Eq.(26)+(27)
\begin{eqnarray}
g_{\rho^-} & =
  & \langle \rho^-|V_1^{\mu}+{\mathrm{i}}V_2^{\mu}|0\rangle\\
{\rm and}\quad
g_{\rho^0} & = & \langle\rho^0|V_3^{\mu}|0\rangle\, .
\end{eqnarray}
They are related by Clebsch--Gordon coefficients
%Eq.(28)
\begin{equation}
\frac{g_-}{g_0} = \frac{\frac{\sqrt{2}}{\sqrt{3}}}
  {\frac{-1}{\sqrt{3}}} = -\sqrt{2}\, .
\end{equation}
The complete coupling of $\rho^0$ to neutrinos through the $Z$--boson
is
%Eq.(29)
\begin{equation}
xg_0 = -\frac{1}{\sqrt{2}}(1-2\sin^2\theta_{\mathrm{W}})\, g_{\rho^-} =
-0.010\quad{\rm for}\quad
 \sin^2\theta_{\mathrm{W}}\approx 0.223\, .
\end{equation}
The resonant $\bar{\nu}\nu$ annihilation cross section, mediated by
hadronic vector mesons is eight times smaller than the corresponding
reaction with electrons as targets, i.e.\ $\sigma_{\bar{\nu}\nu}\sim
0.5\times10^{-38}{\rm cm}^3$. This process contributes beyond the GZK
cutoff only when the mass for one of the relic neutrinos is very small,
$m_{\nu}<10^{-3}\eV$.  In comparison to the Weiler process, the new
cross sections are six order of magnitude smaller.

Multiparticle decays with baryons in the final state are impossible for
$\rho-$mesons. The main decay channel $\rho^0\to \pi^+\pi^-$ has no
observable consequences, since the produced pions decay within
$10^{-6}\pc$ to muons and neutrinos of energy about $10^{16}\eV$, which
are impossible to detect as specific events. The rare decays
$\rho^0\to\pi^0\ga$ and $\rho^0\to\eta \ga$ (different channels are
summarized in Table~\ref{decays}) produce  3 to 7 high--energy photons,
with an opening angle between them being $\sim 10^{-9}$ rad.

\begin{table}[htb]
\caption{Chains of $\rho^0$ meson decays and the corresponding
probabilities.}
\[
\begin{array}{llllll}
\hline
\rho\to \pi^0\ga & 6.8\cdot 10^{-4} & & &
 \pi^0\to 2\ga & 1
\\
\rho\to \eta \gamma & 2.4\cdot 10^{-4} &  &  &
 \eta\to 2\ga & 0.393
\\
& & \eta\to 3\pi^0 & 0.322 & 3\pi^0\to 6\ga & 1
\\
& & \eta\to \pi^+\pi^-\pi^0 & 0.283 & \pi^0\to 2\ga & 1
\\
\hline
\end{array}
\]
\label{decays}
\end{table}

The probabilities are summarized below in Table~\ref{prob}. Strictly
speaking, the resonant cross section is calculated for two pions in the
final state; however, there is no reason to believe that it differs
significantly for other $\rho-$decays. If the annihilation events occur
within a $0.2\pc$ distance from the earth, then these photons can be
observed on the earth as simultaneously arriving $\ga-$initiated
showers.  The number of showers is three or seven if the decaying
$\rho$--meson moves toward the earth and fewer if it moves in another
direction.

\begin{table}[htb]
\caption{Characteristic values for cosmological
$Z$ and $\rho$ bursts.}
\smallskip
\begin{tabular}{p{86mm}@{$\qquad$}cc}
\hline
 & $Z$--burst & $\rho-$burst
\\ \hline
resonant neutrino energy for $m_e=1\eV$, $E_\nu$ &
   $4\cdot 10^{21}\eV$ &  $3\cdot 10^{17}\eV$
\\  %\hline
``one-width'' energy interval, $\Delta E_\nu$ &
      $4.5\cdot 10^{20}\eV$ &  $2.3\cdot 10^{17}\eV$
\\  %\hline
energy--averaged annihilation cross section $\si_{av}$ & $4.2\cdot
10^{-32}\cm^2$ & $0.3\cdot 10^{-38}\cm^2$
\\  %\hline
differential atmospheric neutrino flux at resonant
neutrino energy if it
satisfies the $E_\nu^{-3}$ behaviour estimated in
\cite{volkova80}, $j$ &
$j_0$ & $2.5\cdot 10^{12}\cdot j_0$
\\  %\hline
angle of lab-frame cone & $\sim 10^{-11}$ &
$\sim 10^{-9}$
\\ %\hline
the probability to be directed towards the earth, $w_{\mathrm{earth}}$
& $(\sim 10^{-11})^2$ & $(\sim 10^{-9})^2$
\\ %\hline
distance to observe multishowers, $l_{\mathrm{earth}}$ & $\sim 20\pc$ &
$\sim 0.2\pc$
\\  %\hline
probability of multishower per one annihilation act,
$w_{\mathrm{decay}}$ & &
\\  %
hadronic & $0.7$ &
\\ %
 $3\gamma$ &  & $8.4\cdot 10^{-4}$
\\ %
 $7\gamma$ &  & $7.8\cdot 10^{-5}$
\\
\hline
\end{tabular}
\label{prob}
\end{table}

To estimate the relative probability to observe a multiparticle shower
for different ``bursts'' it is enough to compare the two counting rates
%Eq.(30)
\begin{equation}
P=\si \cdot \Delta E_\nu \cdot j \cdot w_{\mathrm{earth}} \cdot
l_{\mathrm{earth}}^3 \cdot w_{\mathrm{decay}}\, ,
\label{eq30}
\end{equation}
with the variables and their values in Eq. (\ref{eq30})
%$w_{\mathrm{earth}}$ and $w_{\mathrm{decay}}$
being specified in Table~\ref{prob}.  We find
%Eq.(31)
\begin{equation}
\frac{P(\rho)}{P(Z)}\sim 0.1\cdot 10^{-2}
\end{equation}
Thus the probability to observe a signal of $\rho$ or
$K^*$ bursts is even smaller than that for $Z-$bursts.

Other light neutral vector mesons with $J^{PC}=1^{--}$ series are the
$\omega$, whose main decay channel is $\omega \to \pi^+ \pi^- \pi^0$,
and the $\phi$, whose main decay channel is $\phi\to K^0_L K^0_S$ with
the subsequent decays of $K_L^0$ and $K_S^0$ containing several photons
in the final state. The resonant neutrino annihilation through these
mesons also contributes to the events under discussion; but the
$Z^0-$boson coupling to these mesons are unknown.

Considering the $K^{*0}$ meson as the intermediate state, its
contribution to the resonant processes is negligible, because the
$Z^0-$boson coupling to this meson is suppressed, since it involves a
flavour--changing--neutral coupling. Heavy neutral mesons $\rho(1450)$,
$\omega(1420)$, $\phi(1680)$, $\rho(1700)$, $\omega(1650)$ can also
contribute to the resonant annihilation, but at the moment the values
of their couplings to $Z^0$ are hard to estimate.

%Sect.4
\section{Conclusions.}
The cross section of $\bar\nu_l l^-$ and $\nu_l l^+$ reactions have a
resonant behaviour in $s-$channel since the intermediate vector bosons
couple, through vector meson dominance, to mesons with quantum numbers
$J^{PC}=1^{--}$, in particular with the $\rho(770)$ and $K^*(892)$. The
resonant cross sections of reactions $\bar\nu_e e^-\to\rho^-\to
\pi^-\pi^0$ and $\bar\nu_e e^- \to K^{*-}\to (K\pi)^-$ are calculated
to be $4.4\cdot 10^{-38}\cm^2$ and $1.0\cdot 10^{-38}\cm^2$
respectively, and the resonant energies are $580\GeV$ and $780\GeV$.

The $\bar\nu_e e^-$ reaction can be used for the experimental detection
of the electron antineutrinos. In particular, the flux of atmospheric
neutrinos with energies in the region of $\sim 200\GeV$ to $\sim
2000\GeV$, may be visible at neutrino telescopes using water Cherenkov
detectors.

A similar process of resonant annihilation can occur in the universe. A
high energy neutrino of any flavour can annihilate with the
corresponding relic antineutrino, and vice versa, with neutral vector
mesons in the intermediate state. If at least one of the neutrino
species has a mass $\le 10^{-3}\eV$, then the resonant neutrino energy
is $\ge 3\cdot 10^{20}\eV$. These events can lead to multiple the
$\ga-$primaries in cosmic rays beyond the GZK--cutoff. The
neutrino--antineutrino resonant annihilation could also serve as an
indication for the existence of relic neutrinos in the universe. The
probability to observe some signatures on the earth is lower than that
for the Weiler process \cite{weiler99}.

\section*{Acknowledgment}
One of us (O.L.) is grateful to the DAAD for
financial support and the Physics Department of the
University of Dortmund for its hospitality where this
work was performed.  In addition, we thank Prof.\
 G.\ Vereshkov for helpful discussions.
The financial support of the Bundesministerium f\"ur
Bildung
und Forschung under grant No. 05HT1PEA9 is gratefully
acknowledged.

\end{document}